\documentclass[prd,aps,superscriptaddress,twocolumn,floatfix,superscriptaddressno,nofootinbib]{revtex4-1}

\usepackage{graphicx}
\usepackage{indentfirst}
\usepackage{latexsym}
\usepackage{multirow}
\usepackage{tabls}
\usepackage{epsfig}
\usepackage{multirow}
\usepackage{subfigure}
\usepackage{comment}
\usepackage{hyperref}
\usepackage{color}
\usepackage[usenames,dvipsnames]{xcolor}
\usepackage{amssymb}
\usepackage{amsfonts}
\usepackage{amsmath}
\usepackage{bm}
\usepackage{mathrsfs}
\usepackage{dsfont} 
\usepackage{gensymb}
\usepackage[normalem]{ulem}

\hypersetup{colorlinks=true}
\hypersetup{colorlinks,allcolors=black}
\definecolor{MyBlue}{rgb}{0.15,0.15,0.70}
\definecolor{Dgreen}{rgb}{0,0.7,0.0}

\hypersetup{
colorlinks=true,
citecolor=MyBlue,
linkcolor=MyBlue,
urlcolor=MyBlue
}

\newcommand{\be}{\begin{equation}}
\newcommand{\ee}{\end{equation}}

\newcommand{\tobs}{T_\text{obs}}

\begin{document}

\title{Constraining the evolution of Newton's constant with slow inspirals observed from spaceborne gravitational-wave detectors}

\author{Riccardo Barbieri}
\affiliation{Max Planck Institute for Gravitational Physics (Albert Einstein Institute) Am M\"{u}hlenberg 1, 14476 Potsdam, Germany}

\author{Stefano Savastano}
\affiliation{Max Planck Institute for Gravitational Physics (Albert Einstein Institute) Am M\"{u}hlenberg 1, 14476 Potsdam, Germany}

\author{Lorenzo Speri}
\affiliation{Max Planck Institute for Gravitational Physics (Albert Einstein Institute) Am M\"{u}hlenberg 1, 14476 Potsdam, Germany}

\author{Andrea Antonelli}
\affiliation{Department of Physics and Astronomy, Johns Hopkins University, 3400 N. Charles Street, Baltimore, Maryland, 21218, USA}

\author{Laura Sberna}
\affiliation{Max Planck Institute for Gravitational Physics (Albert Einstein Institute) Am M\"{u}hlenberg 1, 14476 Potsdam, Germany}

\author{Ollie Burke}
\affiliation{Max Planck Institute for Gravitational Physics (Albert Einstein Institute) Am M\"{u}hlenberg 1, 14476 Potsdam, Germany}
\affiliation{Laboratoire des 2 Infinis - Toulouse (L2IT-IN2P3), CNRS, UPS, F-31062 Toulouse Cedex 9, France}

\author{Jonathan Gair}
\affiliation{Max Planck Institute for Gravitational Physics (Albert Einstein Institute) Am M\"{u}hlenberg 1, 14476 Potsdam, Germany}

\author{Nicola Tamanini}
\affiliation{Laboratoire des 2 Infinis - Toulouse (L2IT-IN2P3), CNRS, UPS, F-31062 Toulouse Cedex 9, France}

\date\today

\begin{abstract}
    Spaceborne gravitational-wave (GW) detectors observing at milli-Hz and deci-Hz frequencies are expected to detect large numbers of quasi-monochromatic signals.
	The first and second time-derivative of the GW frequency ($\dot f_0$ and $\ddot f_0$) can be measured for the most favourable sources and used to look for negative post-Newtonian corrections, which can be induced by the source's environment or modifications of general relativity.
	We present an analytical, Fisher-matrix-based approach to estimate how precisely such corrections can be constrained.
	We use this method to estimate the bounds attainable on the time evolution of the gravitational constant $G(t)$ with different classes of quasi-monochromatic sources observable with LISA and DECIGO, two representative spaceborne detectors for 
	milli-Hz and deci-Hz GW frequencies.
	We find that the most constraining source among a simulated population of LISA galactic binaries could yield $\dot G/G_0 \lesssim 10^{-6}\text{yr}^{-1}$, while the best currently known verification binary will reach $\dot G/G_0 \lesssim 10^{-4}\text{yr}^{-1}$.
	We also perform Monte-Carlo simulations using quasi-monochromatic waveforms to check the validity of our Fisher-matrix approach, as well as inspiralling waveforms to analyse binaries that do not satisfy the quasi-monochromatic assumption.
	We find that our analytical Fisher matrix produces good order-of-magnitude constraints even for sources well beyond its regime of validity.
	Monte-Carlo investigations also show that chirping stellar-mass compact binaries detected by DECIGO-like detectors at cosmological distances of tens of Mpc can yield constraints as tight as $\dot G/G_0 \lesssim 10^{-11}\text{yr}^{-1}$. 
\end{abstract}
\maketitle

\section{Introduction}
\label{sec:intro}

One of the cornerstones of general relativity is the principle of local position invariance, according to which the outcome of a local nongravitational experiment is independent of the experiment's position in time and space \cite{Will:2014kxa}. Alternatives to general relativity, on the other hand, can violate local position invariance: scalar-tensor theories, for instance, introduce a new field that mediates gravitational interactions~\cite{jordan,Brans:1961sx} and can lead to a time dependence in the effective gravitational constant $G(t)$ \cite{Damour:1988zz,Damour:1992we} replacing Newton's constant $G_0$.
Probing whether the gravitational constant is indeed constant constitutes a direct test of one of the fundamental principles of general relativity, which could potentially provide new insights on the underlying properties of the gravitational interaction at different temporal and spatial scales.

Several strategies have been used over the years to measure the first time derivative of the gravitational coupling, $\dot G$, assuming a linear dependence of $G(t)$ on time (see, e.g., Ref.~\cite{Uzan:2010pm}).
Stringent bounds on $\dot G$ come from Big Bang Nucleosynthesis (BBN) data~\cite{Copi:2003xd,Bambi:2005fi,Alvey:2019ctk}, from which it has been estimated that $\dot G/G_0 \lesssim 10^{-12}\text{yr}^{-1}$, from the Cosmic Microwave Background {(CMB)}~\cite{Wu:2009zb}, and from Type-IA Supernovae~\cite{Gaztanaga:2001fh}. In the local environment, constraints come from the study of globular clusters~\cite{DeglInnocenti:1995hbi} and (at distances below $\sim$1 AU) from lunar ranging experiments, which currently provide the most stringent bounds: $\dot G/G_0  \lesssim 10^{-14}\text{yr}^{-1}$~\cite{Genova}. 

Gravitational-wave (GW) observations have also been used to constrain the running of the gravitational constant. 
{A key difference from previous constraints}
is that GW observations can probe intermediate epochs in cosmic history and are more local in time and space, while cosmological bounds need to assume that the gravitational constant evolved at a constant rate across the entire history of the Universe.
In other words, GWs test the first derivative of $G(t)$ at the spatial and temporal location of the source, without requiring any assumptions on the general form of $G(t)$ at other times.
On the other hand, cosmological measurements of $\dot{G}$ (such as CMB or BBN analyses) assume that $G(t)$ is varying linearly in time from the early universe until today. Thus, GW tests of $\dot{G}$ represent a unique way to probe the local variation of $G(t)$ at cosmological distances.

\begin{figure*}[t]
    \centering
  \includegraphics[width=\textwidth]{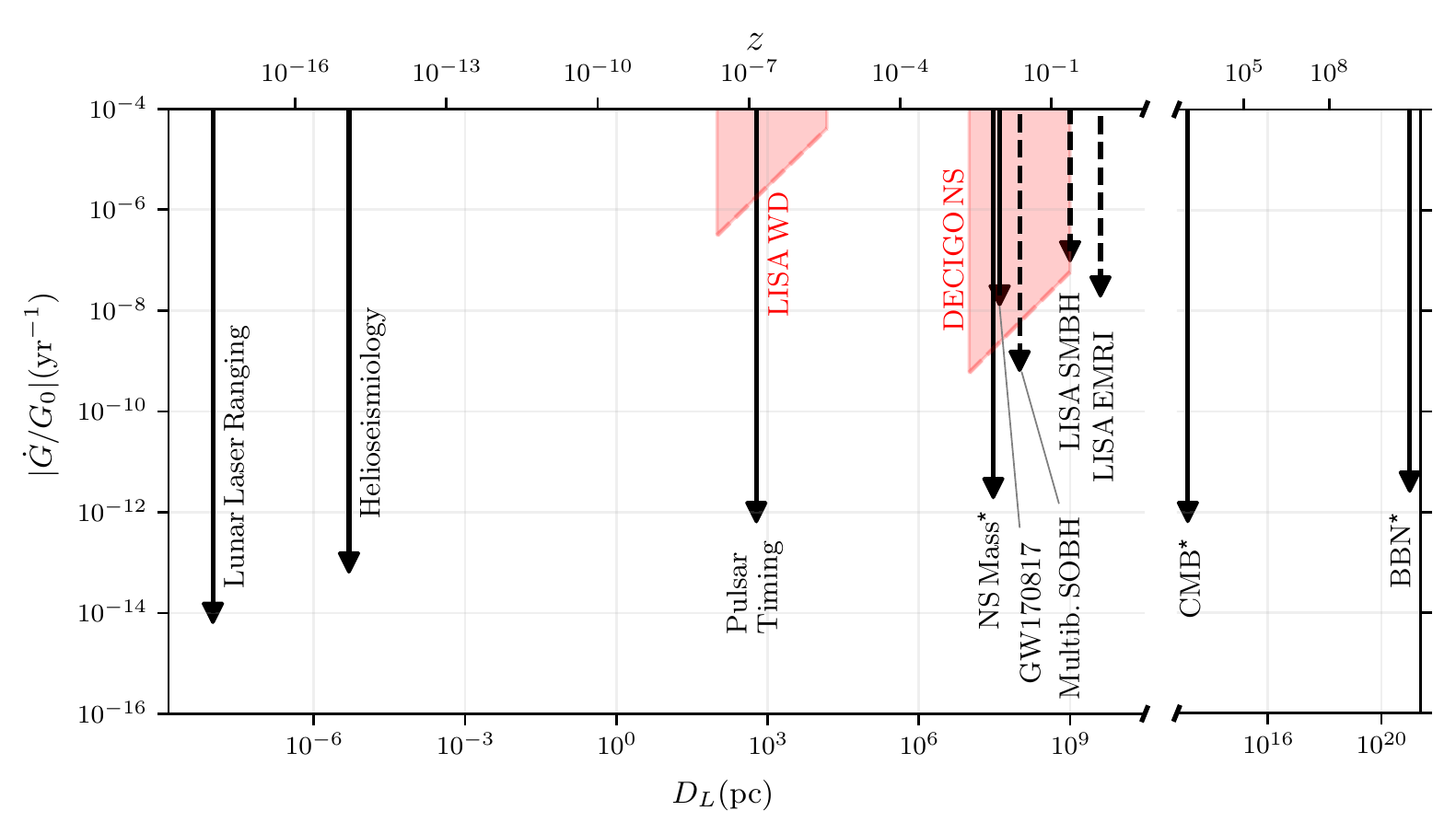}
    \caption{
    Comparison of the constraints on $\dot{G}/G_0$ from current data \cite{Alvey:2019ctk,Wu:2009zb,Genova,Guenther_1998,Kaspi:1994hp,Vijaykumar:2020nzc} (in black, full line) together with predictions for future GW detections from \cite{yunes2009post,Perkins:2020tra} (black, dashed) and predictions from this paper (red bands). The CMB, BBN and NS mass reported constraints assume a linear evolution of $G(t)$ across all cosmological time and their localization in redshift/distance is only approximate (for these reasons we mark them with an asterisk). The estimates from GWs use as reference typical parameters of their respective sources as reported in \cite{yunes2009post,Perkins:2020tra}. These parameters include the value of the distance used in the figure, but the reader should keep in mind that GW constraints can in fact come at different distances and should more properly be represented as a distance/redshift band, similar to the predictions from this paper (in red). The horizontal axes show the distance/redshift (assuming a $\Lambda\text{CDM}$ cosmology with $\text{H}_0=70 \text{km/s/Mpc}$ and $\Omega_m=0.3$) of the sources used in the constraints, while the vertical axis shows the constraint on the magnitude of $\dot{G}$, expressed as a fraction of the value of $G_0$ at the present time (in units of $\text{yr}^{-1}$). The LISA WD constraint band is obtained by considering the Fisher Matrix estimate described in Sec.~\ref{sec:results} for an equal mass DWD with total mass $ M = 2 M_\odot$, a starting frequency $f_0 = 0.01$ Hz, and distances ranging from 100 pc to 15 kpc. The DECIGO NS constraint band is similarly obtained by considering a typical equal mass BNS with a total mass of $M = 2.8 M_\odot$, a starting frequency $f_0 = 0.1$ Hz, and distances between 10 Mpc and 1 Gpc. 
    }
    \label{fig:LiteratureConstraints}
\end{figure*}

Binary pulsars were the earliest GW sources used to place constraints on $\dot G$~\cite{Damour:1988zz,Kaspi:1994hp,Thorsett:1996fr}.
Recently, measurements of component masses of binary neutron stars (NS) have also been used to constrain $\dot G/G_0 \lesssim 10^{-8}\text{yr}^{-1}$~\cite{Vijaykumar:2020nzc}.
It has also been estimated that chirping massive black-hole binaries (MBHBs) and extreme mass-ratio inspirals (EMRIs), which are expected to be detected with the planned LISA mission, could be used to constrain $\dot G/G_0 \lesssim \mathcal{O}(10^{-5})\text{yr}^{-1}$ and $\dot G/G_0 \lesssim \mathcal{O}(10^{-8})\text{yr}^{-1}$ respectively, assuming a $10^6 M_\odot$ central BH and optimistic SNRs of 100 for an EMRI with symmetric mass ratio of $10^{-5}$ and 1000 for an equal-mass MBHB~\cite{yunes2010constraining}.
From stellar-mass binaries, another target of spaceborne GW detectors like LISA, Ref.~\cite{Perkins:2020tra} forecasts constraints in the range $\dot G/G_0 \lesssim 10^{-8}\text{yr}^{-1}$ to $ 10^{-10}\text{yr}^{-1}$, when these sources are observed both by LISA and some configurations of terrestrial detectors [see their Fig.(13)].
In Fig.~\ref{fig:LiteratureConstraints} we summarize current constraints on the variation of the gravitational constant (in black), together with predictions for future GW observations (dashed) and predictions from this paper (in red), plotting them against the reference distance at which those constraints have been derived.
From the figure it is clear that GWs offer the best way to test local variations of $G(t)$ at cosmological distances in the late-time universe.

The goal of this paper is to further assess what constraints stellar-mass compact binaries detected by spaceborne GW interferometers can yield on the running of Newton's constant of gravitation.
The LISA mission is expected to detect tens of thousands of stellar-mass binaries from our Galaxy at milli-Hz frequencies, mostly double white dwarfs (DWDs) \cite{Nelemans2001,Korol:2017qcx,Lamberts2019,Breivik_2020}. Some galactic binaries, known as verification binaries, are already known to emit in the LISA band, and are guaranteed detections \cite{Kupfer:2018jee,Burdge:2019hgl,Kilic:2021dtv}. 
At large separation, the signals from DWDs are almost monochromatic. For most of these, the first time derivatives of the signal's frequencies can be measured and used to constrain the binary's chirp mass. In favourable conditions, the second derivative of the frequency can also be measured, allowing us to measure relativistic effects such as tides \cite{Benacquista:2011gh,Piro:2011qe,McNeill:2019rct} and putative modifications to the environment of the DWDs or of the underlying theory of gravity. Milli-Hz signals are particularly interesting sources to place constraints on $\dot G$, as  the negative post-Newtonian (PN) corrections that $\dot G$ would induce influence the binaries' motion at the large separations at which these signals will be observed.

{Likewise, spaceborne detectors operating at deci-Hz frequencies are expected to detect thousands of binaries containing NS and/or stellar to intermediate-mass black holes (BHs) \cite{Sedda:2019uro}, which could provide stringent constraints to alternative theories of gravity \cite{Yagi:2009zz}. One example of deci-Hz detector is DECIGO \cite{Kawamura:2011zz,Kawamura:2020pcg}, though its actual deployment remains uncertain. In this paper we consider DECIGO as a representative spaceborne deci-Hz detector.}
For low enough chirp masses and frequencies, DECIGO binaries are quasi-monochromatic, which allows us to treat them similarly to LISA's DWDs. For a larger part of the parameter space, low-mass binaries in DECIGO cannot be treated as quasi-monochromatic anymore, as the chirp becomes a dominant feature of the inspiral and the frequency evolution cannot be ignored. Constraints on $\dot G$ from {deci-Hz detectors'} inspiralling sources have not been estimated in the literature and will be provided here for the first time. 

The paper is organised as follows.
Using a Fisher matrix for quasi-monochromatic signals, we derive in Sec.~\ref{sec:analyticalapproach} an analytic estimate of the error attainable in measurements of $\dot G$ that includes correlations with other signal parameters.
We then use this expression to forecast constraints using LISA and DECIGO's quasi-monochromatic binaries in Sec.~\ref{sec:results}.
We first estimate the lowest possible $\dot G/G_0$ constraints attainable from currently known verification DWDs for LISA \cite{Kupfer:2018jee,Burdge:2019hgl,Kilic:2021dtv}, and then survey the parameter space of low-chirp mass galactic binaries in LISA, 
{We further estimate} the constraints from a population study with realistic DWD catalogues. 
{We finally survery} the parameter space of almost-monochromatic binaries in DECIGO.
We find that the loudest known LISA verification binary (ZTF J1539+5027) can be used to constraint $\dot G/G_{0} \lesssim 10^{-4}\text{yr}^{-1}$, while the loudest sources in the simulated DWD population improve this constraint to $\dot G/G_{0} \lesssim 10^{-6}\text{yr}^{-1}$ (thanks to their higher SNR).
In Sec.~\ref{sec:MCMC}, we perform full Bayesian analyses and employ chirping waveforms to explore the parameter space where the quasi-monochromatic approximation fails. We find that LISA could bring constraints down to ${\dot G/G_0 \lesssim 10^{-11}\text{yr}^{-1}}$, if we were to observe stellar-mass BBHs in our Galaxy emitting at the upper end of its sensitivity band.
Likewise, DECIGO could use chirping stellar-mass binaries at cosmological distances to constrain $\dot G/G_0 \lesssim 10^{-11}\text{yr}^{-1}$.
We discuss our results and other prospects in Sec.\ref{sec:discussion}.


\section{Precision measurements of $\dot G$: an analytical approach}
\label{sec:analyticalapproach}

In this section we present an original approach to derive an analytical expression for the constraints on $\dot G$ for almost monochromatic GW sources.
This expression will then be used in Sec.~\ref{sec:results} to assess the potential of LISA and DECIGO to bound $\dot G$.

\subsection{The analytic Fisher matrix}
\label{sec:analytic_FM}

The data stream $d$ observed by a GW detector is assumed to be a superposition of weakly stationary zero-mean Gaussian noise $n(t)$, intrinsic to the detector, and a GW signal $h(t;\boldsymbol{\theta})$ with parameters $\boldsymbol{\theta} = \{\theta_1,\theta_2, ...\}$, 
\begin{equation}\label{eq:data}
    d = h(t;\boldsymbol{\theta}) + n(t).
\end{equation}
Stationary Gaussian noise implies the likelihood is~\cite{whittle1951hypothesis}
\begin{equation}\label{eq:whittle_likelihood}
    \log p(d|\boldsymbol{\theta}) = -\frac{1}{2}(d - h(t;\boldsymbol{\theta})|d - h(t;\boldsymbol{\theta})),
\end{equation}
with inner product in the Fourier domain defined as~\cite{Finn:1992wt}
\begin{equation}\label{eq:inn_prod}
(a|b) = 2\int_{0}^{\infty}\frac{\hat{a}(f)\hat{b}^{\star}(f) + \hat{a}^{\star}(f)\hat{b}(f)}{S_{n}(f)}\text{d}f,
\end{equation}
where $S_n(f)$ is the detector's one-sided Power Spectral Density (PSD)~\cite{khintchine1934korrelationstheorie,wiener1930generalized}, and hatted quantities stand for the continuous Fourier transform. From this, one obtains the (optimal matched-filtering) signal-to-noise ratio (SNR) and the Fisher matrix as, respectively, $\rho=\sqrt{(h|h)}$ and $\Gamma_{ij}=(\partial_i h| \partial_j h)$ (with $\partial_i \equiv \partial/\partial \theta^i$). This latter quantity is of particular interest since it provides an indication of how well parameters can be measured.

In this work, we mostly consider time-domain signals that are quasi-monochromatic (namely, whose frequencies evolve slowly in time). We model these with a sinusoid
\begin{align}\label{eq:GB_signal}
h(t;\boldsymbol{\theta})= A\cos(2\pi t \tilde f
+ \phi)\,.
\end{align}
For compactness, we have gathered the Taylor expansion of the phase into $\tilde f \equiv f_0 + \dot f_0 t/2 +\ddot f_0 t^2/6+\mathcal{O}(\dddot{f}_0^3)$. This is a good approximation as long as ${\dddot f_0 T_{\rm obs}\ll  \ddot f_0}$, with $\tobs$ the observation time of the signal. We take the signal to depend on parameters $\boldsymbol \theta =\{\ln A, f_0, \dot f_0, \ddot{f}_0,\phi\}$.
Following Seto and Takahashi~\cite{Seto:2002dz,Takahashi:2002ky}, we define the (time-domain) inner product for quasi-monochromatic sources as 
\begin{equation}\label{eq:SNR_parseval}
(a|b)|_\text{TD} = \frac{2}{S_{n}(f_{0})}\int_{0}^{\tobs}a(t)b(t)\text{d}t,
\end{equation}
with $f_{0}$ the starting frequency bin. 
The PSD can be moved out of the integral as the PSD in Eq.~\eqref{eq:SNR_parseval} is essentially constant across the frequencies spanned by the evolution of quasi-monochromatic sources. 
With these definitions, the expressions for the SNR and Fisher matrix become 
\begin{align}
\rho^{2} &= \frac{2}{S_{n}(f_{0})} \int_{0}^{T_{\text{obs}}} h(t;\boldsymbol{\theta})^{2} \text{d}t \, ,  \label{eq:analytic_SNR} \\
    \Gamma_{ij} &= \frac{2}{S_{n}(f_{0})} \int_{0}^{T_{\text{obs}}} \partial_{i}h(t;\boldsymbol{\theta})\partial_{j}h(t;\boldsymbol{\theta}) \text{d}t. \label{eq:analytic_fish}
\end{align}
Using equations \eqref{eq:GB_signal} and \eqref{eq:analytic_SNR}, an expression for the SNR can be obtained
\begin{equation}\label{eq:rhoA2}
\rho^2  = \frac{2A^2}{S_n(f_{0})}\int^{\tobs}_0\cos^2\big(2\pi\tilde f t+\phi\big)\text{d}t\approx \frac{A^2\tobs}{S_n}\,,
\end{equation}
where we assumed sufficiently long observation times $f_0\tobs\gg 1$~\cite{Takahashi:2002ky}.
Equation~\eqref{eq:rhoA2} provides a relation between the amplitude $A$ and SNR $\rho$ of the signal.
Using this relation and performing the integrations, use of equation \eqref{eq:analytic_fish} yields the Fisher matrix for the signal up to $\mathcal{O}(\ddot f)$~\cite{Robson:2018svj}
\begin{equation}\label{Gammatri}
\Gamma\approx \rho^2\left(
\begin{array}{ccccc}
1 & 0 & 0 & 0 & 0 \\
0 & \frac{4 \pi ^2 \tobs^2}{3} & \frac{\pi ^2 \tobs^3}{2} & \frac{2 \pi ^2
	\tobs^4}{15} & \pi  \tobs \\
0 & \frac{\pi ^2 \tobs^3}{2} & \frac{\pi ^2 \tobs^4}{5} & \frac{\pi ^2
	\tobs^5}{18} & \frac{ \pi  \tobs^2}{3} \\
0 & \frac{2 \pi ^2 \tobs^4}{15} & \frac{\pi ^2 \tobs^5}{18} & \frac{\pi
	^2 \tobs^6}{63} & \frac{\pi  \tobs^3}{12} \\
0 & \pi  \tobs & \frac{ \pi  \tobs^2}{3} & \frac{\pi  \tobs^3}{12} & 1 \\
\end{array}
\right).
\end{equation}
The measurement precision on the parameter $\theta_i$ is then given by the square root of the diagonal elements of the covariance matrix, namely the inverse of the Fisher matrix:  $\Delta\theta_i = \sqrt{(\Gamma^{-1})_{ii}}$.

In general relativity and in vacuum, measuring the chirp $\dot f_0$ of quasi-monochromatic binaries allows one to break the degeneracy between distance and chirp mass of the source, and it thus gives access to physical parameters of interest~\cite{schutz1986determining}. 
The measurement of $\ddot{f}_0$, while comparatively harder to obtain for quasi-monochromatic sources, would give access to potential tidal interactions in the binary~\cite{Burdge:2019hgl}. If gravity is described by an alternative theory, or if we take into account the binary's environment, these effects can also be accessed through the measurement of $\ddot f_0$, assuming they dominate over potential tidal interactions.

Let us start with a general derivation on how well an additional parameter can be constrained from measurements of $\dot f_0$ and $\ddot f_0$.
Consider the case in which the chirp $\dot f_0$ and second derivative $\ddot f_0$ depend on two parameters $\theta_1$ (such as the chirp mass) and $\theta_2$ (any modification to GR or the binary's environment). Using the chain rule, we can swap the $\dot f_0$ and $\ddot f_0$ entries in the Fisher matrix $\Gamma$ with $\theta_1$ and $\theta_2$. The new Fisher matrix $\tilde\Gamma$ has parameters $\theta=\{\log A,f_0,\theta_1,\theta_2,\phi\}$ and is given by
\begin{equation}\label{eq:FM_new}
\tilde\Gamma=J_\theta^T\, \Gamma\, J_\theta\,,
\end{equation} 
where $(\cdot)^T$ denotes the transpose operation and $J_\theta$ is the Jacobian 
\begin{equation}\label{Jtheta}
J_\theta=\begin{pmatrix}
1 & 0 & 0 & 0 & 0 \\
0 & 1 & 0 & 0 & 0\\
\frac{\partial \dot f_0}{\partial \ln A}&\frac{\partial \dot f_0}{\partial f_0}&\frac{\partial \dot f_0}{\partial \theta_1}&\frac{\partial \dot f_0}{\partial \theta_2}&\frac{\partial \dot f_0}{\partial \phi}\\
\frac{\partial \ddot f_0}{\partial \ln A}&\frac{\partial \ddot f_0}{\partial f_0}&\frac{\partial \ddot f_0}{\partial \theta_1}&\frac{\partial \ddot f_0}{\partial \theta_2}&\frac{\partial \ddot f_0}{\partial \phi}\\
0 & 0 & 0 & 0 & 1
\end{pmatrix}\,.
\end{equation}
The estimates of measurement precision, $\Delta\theta_1$ and $\Delta\theta_2$, can then be obtained by inverting $\tilde\Gamma$ and reading off the diagonal elements. For sufficiently simple first and second chirp derivatives, the expressions are analytical and therefore cheap to evaluate, but still automatically incorporate correlations between parameters.

\subsection{Time-varying gravitational constant: a nonperturbative model for the GW frequency}

We now demonstrate how to use the analytical approach for quasi-monochromatic sources reported above to place constraints on the time variation of the gravitational constant $G(t)$. 

We first expand $G(t) = G_0+ \dot{G} (t-t_0)+\mathcal{O}[(t-t_0)^2]$ about Newton's gravitational constant $G_0$ at the initial time of observation $t_0$, with $\dot{G} \equiv \dot{G}(t_{0})$. Previously, waveforms accounting for the running of Newton's constant had been presented in Ref.~\cite{yunes2010constraining}. That analysis focused on chirping binaries, for which an expansion of $G(t)$ around the time of coalescence of the binary (rather than the initial time) is more appropriate. Waveforms in Ref.~\cite{yunes2010constraining} were also only valid up to leading order in $\dot G$. This approximation breaks down in some of the parameter space we explore: higher orders in $\dot G$ can only be safely neglected at sufficiently high frequencies, for which $\dot G/G_0 \ll  f_0^{8/3} G_0^{5/3} \mathcal{M}_c^{5/3} c^{-5} $ (as seen comparing the first and second terms in  Eq.~\eqref{eq:fdot_and_fddot_withGdot} below).
In this work, we derive quasi-monochromatic waveforms valid to all powers in a constant $\dot G$~in the phase.
   
We start by solving the balance equation $\dot E = - \mathcal{L}_\text{GW}$, with $E$ and $\mathcal{L}_\text{GW}$ the binary's binding energy and GW emission power, respectively, at leading PN order \cite{Maggiore:2007ulw}
\begin{align}
    E &= - \left[\frac{\pi^2}{8}\mathcal{M}_c^5 G(t)^2 f(t)^2\right]^{1/3}, \nonumber\\
    \mathcal{L}_\text{GW} &= \frac{32}{5}\frac{c^5}{G(t)}\left[\frac{\pi \mathcal{M}_c}{c^3}G(t)f(t)\right]^{10/3}.
\end{align}
These determine the evolution of the frequency as a function of time.
Defining $f_0\equiv f(t_0)$, the expressions for $\dot f_0 \equiv \dot f (t_0)$ and $\ddot f_0 \equiv \ddot f (t_0)$ from the balance law are
\begin{align}
    \dot f_0 =& \frac{96\pi^{8/3}}{5 c^5}G_0^{5/3}\mathcal{M}_c^{5/3}f_0^{11/3}  - \left(\frac{\dot G}{G_0}\right)f_0, \nonumber\\
    \ddot f_0 =& \frac{33792\pi ^{16/3}}{25 c^{10}} f_0^{19/3} G_0^{10/3} \mathcal{M}_c^{10/3} \nonumber\\
    &-\frac{288\pi ^{8/3}}{5 c^5} f_0^{11/3} G_0^{5/3} \left(\frac{\dot G}{G_{0}}\right) \mathcal{M}_c^{5/3} +2\left(\frac{\dot G}{G_0}\right)^{2}f_0.
   \label{eq:fdot_and_fddot_withGdot}
\end{align}
From the expression for $\dot f_0$, it is apparent that $\dot G$ formally enters at $-4$PN order, i.e., its correction scales as $f^{2n/3}$ with $n=-4$ relative to the leading-order general relativistic term. This is also true for $\ddot{f}_0 $ and all other higher derivatives of the frequency, assuming an expansion in $\dot G$ to linear order.
Note that with respect to the analysis in Ref.~\cite{yunes2010constraining} we are keeping all terms in Eqs.~\eqref{eq:fdot_and_fddot_withGdot}, in particular the first and last term on the right hand side respectively of the first and second line of Eqs.~\eqref{eq:fdot_and_fddot_withGdot}, since we are not assuming the condition $\dot G/G_0 \ll  f_0^{8/3} G_0^{5/3} \mathcal{M}_c^{5/3} c^{-5} $.

In the LISA band, for instance, for galactic binaries of $\mathcal{M}_c = 0.5 M_\odot$ at frequencies $f_0 = 10^{-2}$Hz the first term in $\dot f_0$ dominates for values $\dot G/G_0 > 10^{-5}\text{ yr}^{-1}$, and must therefore be included. Note that for low-enough frequencies, the terms of $\dot G$ in both $\dot f_0$ and $\ddot f_0$ may imply that $\dddot f_0 T_{\rm obs}>\ddot f_0$, thus breaking the quasi-monochromatic assumption. 
We take into account this limitation in our analysis.

The expressions \eqref{eq:fdot_and_fddot_withGdot} provide $\dot f_0 (\mathcal{M}_c,\dot G)$ and $\ddot f_0 (\mathcal{M}_c,\dot G)$ in terms of two new parameters of interest, the chirp mass $\theta_1\equiv \mathcal{M}_c$ and the parameter $\theta_2\equiv\dot G$: ultimately, we want to extract information about the latter, keeping track of correlations with the former (along with those with $A$, $f_0$ and $\phi$). Using Eqs.~\eqref{Jtheta} and \eqref{eq:fdot_and_fddot_withGdot}, we can recast the Fisher matrix \eqref{eq:FM_new} in terms of the $\mathcal{M}_c$ and $\dot G$ parameters using the results derived in Sec.~\ref{sec:analytic_FM}. The measurement error on $\dot G$ can then be obtained through $\Delta\dot G_\text{full} = \sqrt{(\tilde\Gamma^{-1})_{\dot G\dot G}}$. The expression we obtain is rather cumbersome, but it can be shown to be well approximated by a simpler expression. Defining the quantities
\begin{align}
    \epsilon_1 \equiv & \frac{\dot G}{G_0}\tobs = 0.01 \left( \frac{\dot{G}/G_0}{10^{-2} {\rm yr}^{-1}} \right) \left( \frac{T_{\rm obs}}{1 {\rm yr}} \right), \label{eq:epsilon1}\\
    \epsilon_2 \equiv & \frac{\pi ^{8/3}}{c^5}f_0^{8/3} G_0^{5/3} \mathcal{M}_c^{5/3}\tobs \nonumber \\
    \simeq& 0.01 \left( \frac{f_0}{10^{-2} {\rm Hz}} \right)^{8/3} \left( \frac{\mathcal{M}_{\text{c}}}{10^{2} M_{\odot}} \right)^{5/3} \left( \frac{T_{\rm obs}}{1 {\rm yr}} \right), \label{eq:epsilon2} 
\end{align}
the measurement error reads
\begin{align}\label{eq:DeltaGdot_full}
     \Delta\dot G_\text{full}^2 =& \Delta\dot G_\text{app.}^2
     \bigg[1-3\epsilon_1+\frac{88}{35}\epsilon_1^2-\frac{9}{28}\epsilon_1^3+\frac{1 }{84}\epsilon_1^4\\
     &+\frac{704}{5}\epsilon_2-\frac{42944}{175}\epsilon_1\epsilon_2+\frac{2112 }{35}\epsilon_1^2\epsilon_2-\frac{352}{105}\epsilon_1^3\epsilon_2\nonumber\\
     &+\frac{4697088}{875}\epsilon_2^2-\frac{447744}{175}\epsilon_1\epsilon_2^2+\frac{5632 }{21}\epsilon_1^2\epsilon_2^2\nonumber\\
     &+\frac{17842176}{875}\epsilon_2^3-\frac{3964928}{875}\epsilon_1\epsilon_2^3+\frac{95158272
   }{4375}\epsilon_2^4\bigg],\nonumber
\end{align}
where 
\begin{equation}\label{eq:deltaGdot3}
 \Delta\dot G_\text{app.}^2 = \frac{630000\  c^{10} G_0^4}{\pi ^2 f_0^2 \tobs^6 \rho^2 \left(5 c^5 \dot G+416 \pi ^{8/3} f_0^{8/3} G_0^{8/3}
   \mathcal{M}_c^{5/3}\right)^2}.
\end{equation}
Since $\epsilon_1$ and $\epsilon_2$ remain generally small for the sources we are interested in -- see values in Eqs.~\eqref{eq:epsilon1}, \eqref{eq:epsilon2} 
-- we find that $\Delta\dot G_\text{app.}$ is a good and simple approximation to estimate the error on $\dot G$. However, we use $\Delta\dot G_\text{full}$ in the analytical estimates below for completeness.

\section{Constraints on $\dot{G}$ from the analytic Fisher matrix}
\label{sec:results}

By using the analytical expression Eq.~\eqref{eq:DeltaGdot_full}, in this section we explore the constraints on $\dot G$ across the whole parameter space of low-mass, quasi-monochromatic binaries detected by LISA and {deci-Hz detectors, using DECIGO as an example.}

Following Cornish~\cite{robson2019construction}, we define the angle-averaged gravitational wave amplitude of a quasi-monochromatic source, 
\be\label{eq:GB_amplitude}
A = \frac{8}{\sqrt{5}} \frac{(G_0 \mathcal{M}_{\text{c}}/c^2)^{5/3}}{D_L} \bigg(\frac{\pi f_0}{c}\bigg)^{2/3}\; ,
\ee
for $D_{L}$ the luminosity distance to the source. For the amplitude in equation \eqref{eq:GB_amplitude}, we approximate $G\simeq G_0$, as GW detectors are in any case less sensitive to modulations of the amplitude compared to modulations of the phase. 
We compute LISA SNRs with the noise PSD from Ref.~\cite{robson2019construction}, and DECIGO SNRs using the noise PSD from Ref.~\cite{Yagi:2011wg}. 

In the present section, we consider $\dot G$ to be measurable if its 1$\sigma$ relative error reaches a 50\% precision or lower~\cite{Robson:2018svj}, $\Delta\dot G_\text{full}/\dot G < 0.5$. For quasi-monochromatic sources, this condition can be analytically evaluated using Eq.~\eqref{eq:DeltaGdot_full}, allowing us to cheaply survey the parameter space.
We will quote measureable values of $\dot G$ corresponding to the minimum $\dot G$ we can detect for a given set of parameters.
Values lower than those quoted would imply a degradation in the precision of the measurement; higher values would imply an even better detection of $\dot G$.

We limit our survey to regions of parameter space where the quasi-monochromatic approximation applies by requiring that the third time derivative of the frequency $\dddot f_0$ is such that it's contribution to the Taylor-expanded frequency evolution is negligible, namely ${\dddot f_0 T_{\rm obs}\ll \ddot f_0}$.
The third time-derivative can easily be found from $\dot f$ and $\ddot f$ in Eq.~\eqref{eq:fdot_and_fddot_withGdot}. 
Constraining the parameter space in this way singles out two source classes of interest: binaries with NSs, BHs or DWDs in our galaxy (observable with LISA), and NS or BH binaries at cosmological distances of tens of Mpc (observable with DECIGO).

\subsection{LISA}
\label{subsec:LISA_analytic}

\begin{table*}[ht]
	\centering
	\caption{Forecast constraints on $\dot G$ from LISA verification (detached) binaries. The values refer to $4.5$ years of continuous observation.
	}
	\label{tab:LISA_ver_binaries}
	\begin{tabular}{|c c | cccc | c| } 
		\hline
		Source & Ref. & $f_0$[mHz] & $m_1[M_\odot]$ & $m_2[M_\odot]$ & $D_L$ [kpc] & $ \dot G/G_0\  [ \text{yr}^{-1}]$ \\
		\hline

        {\bf ZTF J1539+5027} & \cite{Burdge:2020end,littenberg2019prospects} & 4.8 & 0.61 & 0.21 & 2.34 & 2.05 $\times 10^{-4}$\\
        ZTF J0538+1953 & \cite{Burdge:2020end} & 2.3 & 0.45 & 0.32 & 0.68 & 2.76 $\times 10^{-4}$\\
        PTF J0533+0209 & \cite{Burdge:2020end} & 1.6 & 0.65 & 0.17 & 1.74 & 9.56 $\times 10^{-4}$\\
        ZTF J2029+1534 & \cite{Burdge:2020end} & 1.6 & 0.32 & 0.30 & 2.02 & 1.05 $\times 10^{-3}$\\
        ZTF J0722$-$1839 & \cite{Burdge:2020end} & 1.5 & 0.38 & 0.33 & 0.93 & 7.20 $\times 10^{-4}$\\
        ZTF J1749+0924 & \cite{Burdge:2020end} & 1.3 & 0.40 & 0.28 & 1.55 & 1.27 $\times 10^{-3}$\\
        ZTF J2243+5242 & \cite{Burdge:2020bul} & 3.8 & 0.35 & 0.38 & 2.12 & 2.44 $\times 10^{-4}$\\
        SDSS J0651+2844 & \cite{Hermes:2012us} & 2.6 & 0.26 & 0.51 & 1.00 & 2.86 $\times 10^{-4}$\\
        SDSS J0935+4411 & \cite{Kilic:2014fea} & 1.7 & 0.32 & 0.14 & 0.66 & 7.54 $\times 10^{-4}$\\
        SDSS J2322+0509 & \cite{Brown:2020uvh} & 1.7 & 0.27 & 0.24 & 0.76 & 6.85 $\times 10^{-4}$\\
        SDSS J1630+4233 & \cite{Kilic:2011ej}  & 0.8 & 0.30 & 0.30 & 0.70 & 2.41 $\times 10^{-3}$\\
        SDSS J1235+1543 & \cite{Gemini2017}    & 0.6 & 0.35 & 0.17 & 0.39 & 3.61 $\times 10^{-3}$\\
        SDSS J0923+3028 & \cite{Brown:2010sa}  & 0.5 & 0.28 & 0.76 & 0.28 & 2.62 $\times 10^{-3}$\\
		\hline
	\end{tabular}
\end{table*}

\begin{figure*}
\includegraphics[width = 0.43\textwidth]{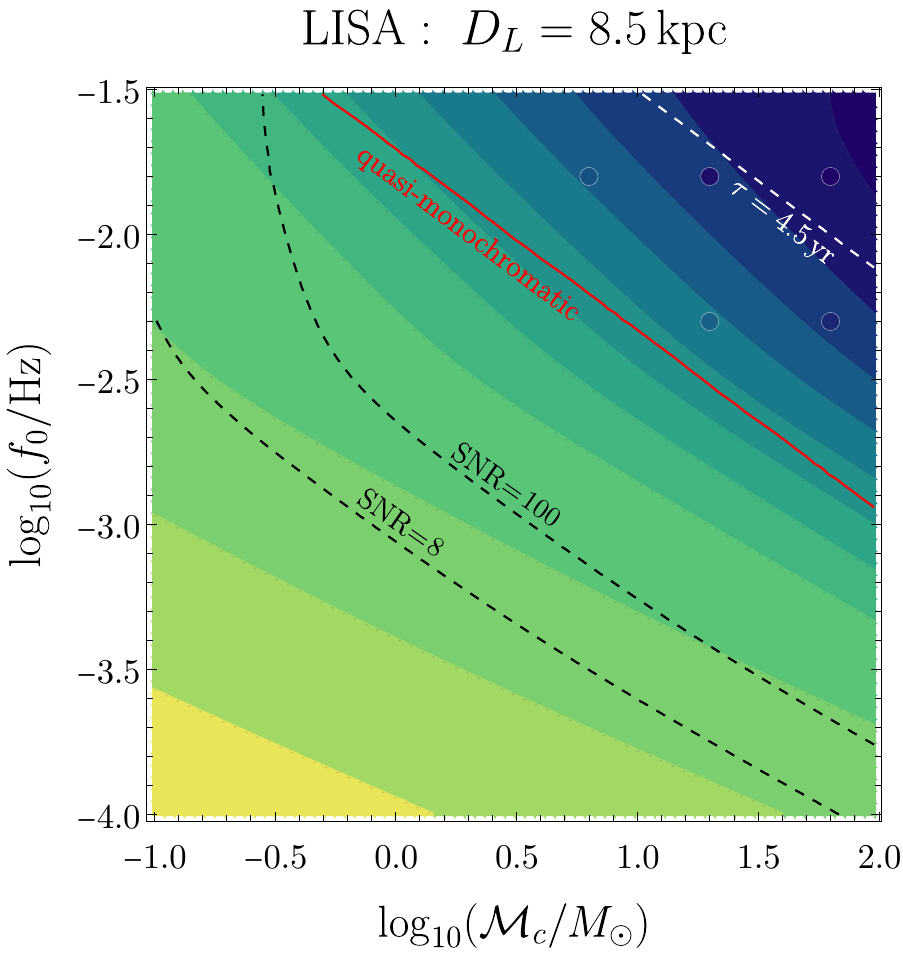} \hspace{0.1cm}
\includegraphics[width = 0.545\textwidth]{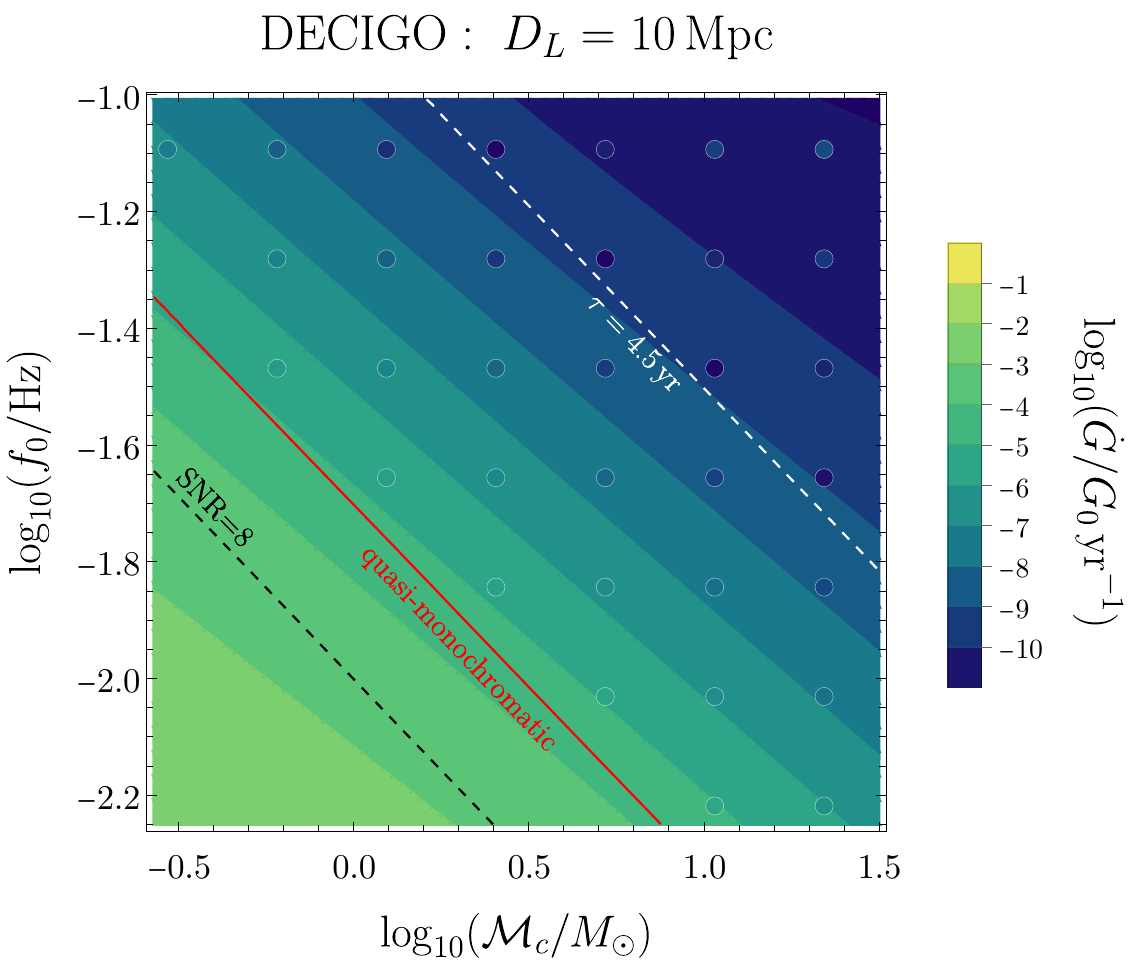}
\caption{The color maps display the smallest $\dot{G}/G_0$ observable with 50\% precision as a function of the chirp mass of the source, $\mathcal{M}_c$, and frequency $f_0$. 
In the regions in the top-right corners above the solid red line, the quasi-monochromatic approximation is not valid. 
In the LISA plot, the quasi-monochromatic approximation also breaks down in the bottom-left corner, where the value of $\dot G$ increases significantly and its effect dominates over radiation backreaction.
Dashed black lines mark SNR levels, while sources above the dashed white lines merge within the observational time of $4.5$ yr.
The coloured dots display bounds sampled through an MCMC analysis using chirping waveforms. 
}
\label{fig:contourGdot}
\end{figure*}

The first sources we consider are ultra-compact binaries in our Galaxy. These are short-period ($P\lesssim 1$ hour) binaries generally composed of white dwarves, NSs and compact helium-stars, which we expect to detect in the tens of thousands with LISA~\cite{Nelemans2001,Korol:2017qcx,Lamberts2019,Breivik_2020}. 
Some of these binaries are so loud that they will be detectable within the first few weeks of mission operation.
Others have already been detected by EM telescopes and will act as verification binaries for the detector's performance.

Using the frequencies, masses and distances quoted in Refs.~\cite{Kupfer:2018jee,Burdge:2019hgl,Kilic:2021dtv}, we can estimate the lowest measurement of $\dot G$ attainable with LISA verification binaries.
We focus on detached binaries, namely those that are not undergoing mass transfer (which would modify the chirp more than any expected effect from $\dot G$), and on systems where the two binary stars are clearly distinguishable in EM observations, implying negligible tidal effects.
The results are reported in Table~\ref{tab:LISA_ver_binaries}, where we assumed $4.5$ years of continuous observations by LISA.
The shortest-period verification binary known to date, ZTF J1539 (with angle-averaged SNR$=138$), yields the most stringent constraint for $\dot{G}$ at $\dot G/G_0\lesssim 10^{-4}\text{yr}^{-1}$. 
In general, currently known verification binaries lead to constraints between $\dot G/G_0\lesssim 10^{-3}\text{yr}^{-1}$ and $10^{-4}\text{yr}^{-1}$.

While the constraints from verification binaries fall short of prospective constraints from chirping massive BH binaries with LISA \cite{yunes2009post,Perkins:2020tra}, these constraints are guaranteed. Moreover, the strength of galactic DWDs lies in numbers. We therefore explore how well a full, realistic population of galactic DWDs will constrain $\dot G$. Analyzing a dataset produced by the population synthesis code \textsc{SeBa} used in~\cite{Portegies1996,Nelemans2001b,Toonen2012} (assuming 4 years of observations), we find that the best constraint LISA could obtain from a single DWD in the population is
\begin{equation}
  \frac{\dot{G}}{G_0} \bigg|_{\rm best \ DWD} \lesssim
 3.0 \times 10^{-6}  \, {\rm yr}^{-1}.
 \label{eq:DWD_best_constr}
\end{equation}
Assuming for simplicity that LISA's DWD observations are independent\footnote{In reality, DWDs will be observed simultaneously in LISA and there could be small correlations between their inferred parameters, which would degrade measurements of other common parameters.}, and assuming that $\dot{G}$ takes the same value across the Galaxy, we can combine all constraints in the population and improve slightly over the best event. Requiring $\Delta\dot G =
\left( \sum_\text{pop.}\Delta\dot G_{\rm full}^{-2}\right)^{-1/2} < 0.5 \, \dot G$, we find 
\begin{equation}
 \frac{\dot{G}}{G_0} \bigg|_{\rm DWD \ pop.} \lesssim 
2.8 \times 10^{-6} \, {\rm yr}^{-1}. 
\end{equation}
This marks a significant improvement from the constraints one can set with verification binaries alone, but only a minor improvement with respect to the most significant binary in the population which yields the constraint~\eqref{eq:DWD_best_constr}.
The constraints attainable with a realistic population of DWDs are in fact competitive with those achievable with massive BH binaries and EMRIs observed by LISA~\cite{yunes2009post,Perkins:2020tra}.

We also perform a parameter-space survey of all galactic sources potentially detectable by LISA, with masses that encompass both DWDs and other more massive populations of nearby compact-object binaries.  In Fig.~\ref{fig:contourGdot} we show results for frequencies and chirp masses within $f_0=[10^{-4},3 \times 10^{-2}]$ Hz and $\mathcal{M}_c =[0.1, 100]\ M_\odot$, respectively. We further assume that all binaries are equal-mass, as the results do not depend on the mass ratio, see Eqs.~(\ref{eq:DeltaGdot_full}-\ref{eq:deltaGdot3}). 
The high-end of the mass spectrum corresponds to 
intermediate-mass BHs that might be present in the Milky Way's globular clusters, see e.g., Ref.~\cite{Strokov:2021mkv} for discussions about this scenario, or Ref.~\cite{Ye:2021lnk} for an example of the potential compact object population of a Galactic globular cluster.
Throughout the parameter space we fix the observation time to $\tobs = 4.5$ yr and the distance to $D_L = 8.5$ kpc, which corresponds to the Milky Way's Galactic center where the majority of DWDs are expected to reside.
The SNR varies with the source's chirp mass, as indicated by the black dashed lines in Fig.~\ref{fig:contourGdot}.

The survey confirms that in the LISA band, for the small masses of Galactic binaries and small values of $\dot G$, most signals are quasi-monochromatic and can constrain values down to $\dot G/G_0\lesssim 10^{-7}\text{yr}^{-1}$. This constraint will be achieved with massive sources at low frequencies (Galactic intermediate-mass BH binaries) or lighter sources at higher frequencies.
The quasi-monochromatic, analytic estimates in the upper right corner of Fig.~\ref{fig:contourGdot} suggest that LISA could achieve even better constraints if it detects stellar-mass BH binaries in our Galaxy. We explore this scenario further in Sec.~\ref{sec:MCMC} using chirping waveforms.

We have also explored the constraints LISA could obtain from extra-Galactic ($D_L=100$ Mpc) stellar- and intermediate-mass BHs. At these distances, however, LISA could provide only modest constraints $\dot G/G_0\lesssim 10^{-2}\text{yr}^{-1}$ for a restricted region of parameter space. We conclude that for more massive binaries at cosmological distances, chirping sources work best to constrain $\dot G$, as argued in Ref.~\cite{yunes2010constraining}.

\subsection{DECIGO}

{Deci-Hz detectors are} sensitive to NS and BH binaries up to redshift $z\sim10$ \cite{Sedda:2019uro,Takahashi:2002ky}. Those observed at large separations may be approximated as quasi-monochromatic signals. We can therefore perform a similar parameter-space survey for these sources. We {investigate sources detected by} DECIGO at a fixed cosmological distance $D_L=10$ Mpc, and assume equal-mass sources and an observation time of $4.5$ yrs. 

In the right panel of Fig.~\ref{fig:contourGdot} we see that, in the DECIGO frequency band, only binaries with chirp masses (and roughly equal component masses) $\lesssim 15 M_\odot$ are quasi-monochromatic. 
{These include} binaries containing low-mass stellar-origin BHs or NSs, or one of each. For these, the best constrains are around $\dot G/G_0\lesssim 10^{-4}\text{yr}^{-1}$. 

Fig.~\ref{fig:contourGdot} shows that, for essentially all source masses, observing the chirping phase 
will be crucial to obtain good constraints on $\dot G$. Note that the rates of NS binaries (the observable quasi-monochromatic sources in the right panel of Fig.~\ref{fig:contourGdot}) so close to us are uncertain; however, even for those that are further away than what is suggested here,
observing the chirping phase will improve bounds by orders of magnitude~\cite{Vijaykumar:2020nzc}.

\section{Full Bayesian analysis}
\label{sec:MCMC}
The predictions obtained with the Fisher matrix formalism are particularly useful to quickly estimate the constraints on $\dot{G}$ over the parameter space. However, this formalism is known to provide a reliable estimate of the measurement precision only in the regime where the linear signal approximation is valid, which requires the SNR to be high. Therefore, in this section we use Markov Chain Monte Carlo (MCMC) methods to sample from the posterior distribution and check the predictions of the Fisher matrix formalism.
For sampling, we use the \texttt{emcee} package \cite{emcee}.

We also want to compare the Fisher Matrix analysis and the full Bayesian analysis in the region of parameter space where the quasi-monochromatic approximation fails, since we also expect that fully chirping binaries provide the tightest constraints. In the full Bayesian analysis, we use either quasi-monchromatic waveforms of the kind defined by Eq.~\eqref{eq:GB_signal} or chirping waveforms, depending on the parameters of the source. The `chirping' waveform we employ has an IMRPhenomD phase \cite{Husa:2015iqa,Khan:2015jqa} modified to include the effect of the running of $G$ to leading order in $\dot G$. The latter is analogous to the leading phase contribution due to mass accretion~\cite{Caputo:2020irr} (replacing the accretion parameter $ f_{\rm Edd}/\tau$ with $\dot G_0/G_0$) or peculiar acceleration~\cite{Tamanini:2019usx}.

\subsection{Quasi-monochromatic LISA sources}
\begin{figure*}[t]
    \centering
    \includegraphics[width=.7\textwidth]{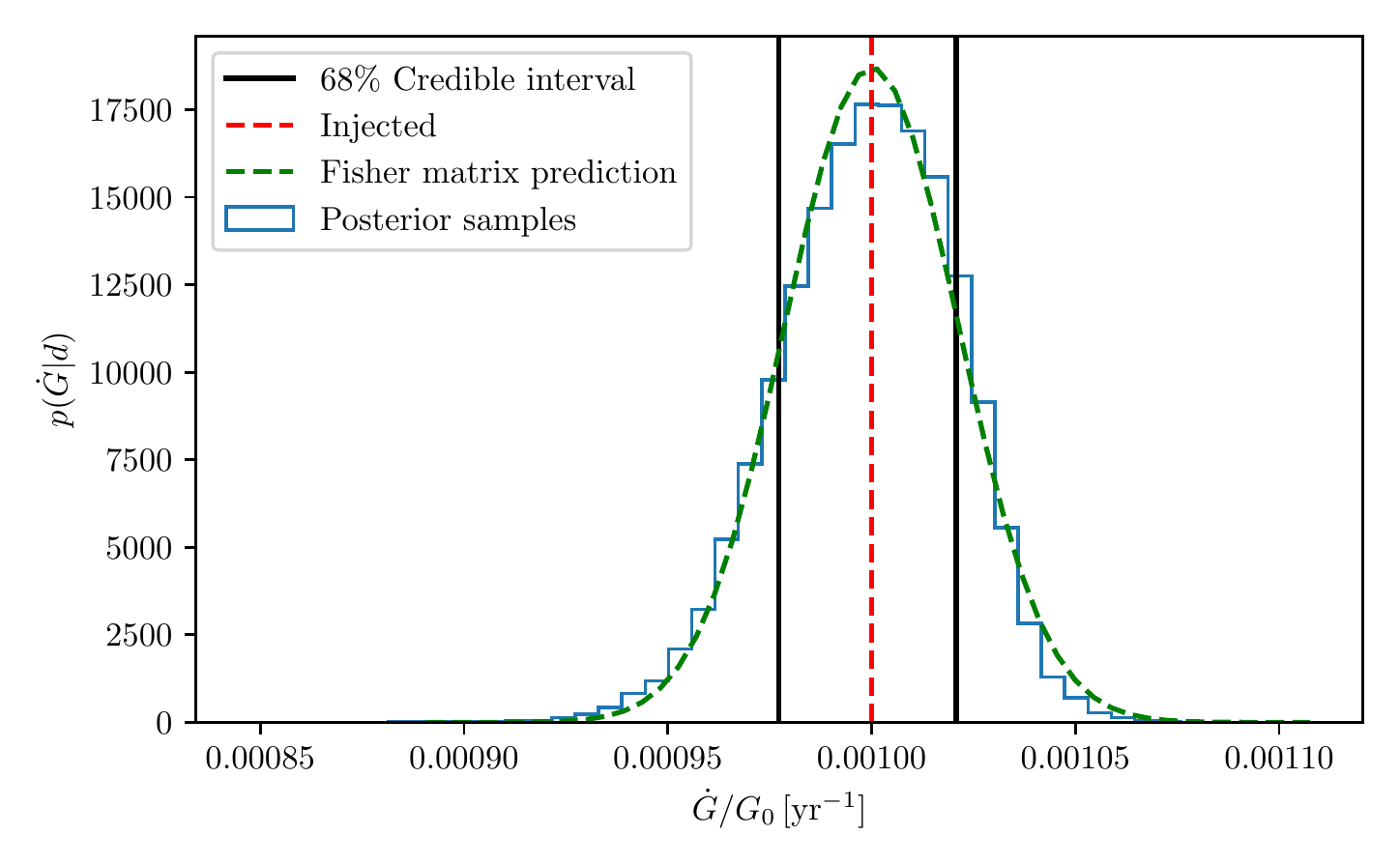}
    \caption{Marginalized posterior distribution (blue) of $\dot{G}$ for a binary with the frequency and distance of ZTF J1539+5027, observed by LISA. For a true value of $\dot{G}/G_0=1 \times 10^{-3}$ yr$^{-1}$ (red), we predict a measurement of $\dot G/G_0= 0.001 ^{+2.3e-5} _{-2.1e-5}$ yr$^{-1}$ (68 \% credible interval, in black). The Fisher matrix prediction for the posterior (dashed green) is in good agreement with the result of MCMC sampling, with the width of the posterior agreeing with the Fisher matrix estimate to within 2\%. }\label{fig:mcmc_gdot}
\end{figure*}

{In this section, we use the quasi-monochromatic waveforms of Eq.~\eqref{eq:GB_signal} in the Bayesian analysis, as these are the ones that can be directly compared to the Fisher estimates, which was also obtained assuming the signal model given in Eq.~\eqref{eq:GB_signal}. 
Here, we check the measurement of $\dot{G}$ for one of the best performing verification binaries, ZTF J1539+5027.
As predicted by the Fisher matrix (see Table \ref{tab:LISA_ver_binaries}), any value above $\dot{G}/G_0>2.05\times10^{-4} \text{yr}^{-1}$ will be measured with a relative precision larger than 50\%. Therefore, we use $\dot{G}/G_0 = 1\times10^{-3} \text{yr}^{-1}$, above the detection limit, and use the frequency, mass and distance parameters of ZTF J1539+5027 \cite{Burdge:2020end,littenberg2019prospects} to inject the signal. The resulting parameter values are: 
$A =1.77\times10^{-22}$, $f_0=4.8\times10^{-3}\text{Hz}$, $\dot f_0= -1.5\times 10^{-13}\text{Hz} \, s^{-1}$, $\ddot f_0=9.7\times 10^{-24}\text{Hz} \, s^{-2}$, $\phi=0.2 \, {\rm rad} $.
}

We first transform the waveform~\eqref{eq:GB_signal} to the Fourier domain with the 1st order stationary phase approximation {(valid for the system we selected)}. 
We sample over the parameters $\boldsymbol \theta =\{ A, f_0, \dot f_0, \ddot{f}_0,\phi \}$ as commonly done in the literature \cite{Littenberg:2020bxy, littenberg2019prospects}. The posterior samples from the MCMC are then converted into posterior samples in $f_0, \dot G, \mathcal{M}_c$ through Eq.~(\ref{eq:fdot_and_fddot_withGdot}).

We show the marginalized posterior for the ZTF-like binary in Fig.~(\ref{fig:mcmc_gdot}). The derivative of Newton's constant is measured as $\dot G/G_0= 0.001 ^{+2.3e-5} _{-2.1e-5}$ yr$^{-1}$. Most importantly, the width of the posterior distribution agrees with the Fisher matrix prediction with a ratio of the two uncertainties approximately equal to $1.02$. This validates the results presented in the previous section, and in particular in Fig.~\ref{fig:contourGdot}, within the limits of validity of the monochromatic approximation. 

\subsection{Chirping LISA sources}

In order to assess the limitations of the quasi-monochromatic approximation, we perform full Bayesian analysis with a chirping waveform model for sources in the upper right corner of the parameter space in Fig.~\ref{fig:contourGdot}. Sources falling in this region of parameter space would be BH or NS binaries residing in our Galaxy and with rapid frequency evolution, see Ref.~\cite{Wagg:2021cst}. 

We aim to find the constraint on $\dot{G}$ using the same technique employed in the Fisher Matrix analysis: find the smallest $\dot{G}$ measurable at 50\% precision at 1$\sigma$. We perform several MCMC runs with different injected values of $\dot G$, and identify as our constraint the value that produces a 1$\sigma$ relative error between 40\% and 50\%. When using the chirping waveform, convergence is easier when sampling on $\mathcal{M}_c$, $\eta$, $f_0$, $\phi$, $D_{L}$ and $\dot G$. 

We find that LISA could actually achieve constraints comparable to the ones predicted by our quasi-monochromatic approximation everywhere in the upper right corner of Fig.~\ref{fig:contourGdot}. The best constraint that we identified was $\dot G / G_0 = 8.5 \times 10^{-12} \text{yr}^{-1}$, achieved with a binary emitting at a frequency $f_0\simeq0.0158 $ Hz and with a chirp mass $\mathcal{M}_{\text{c}}\simeq60 M_{\odot}$.
These results clearly show that the monochromatic analytical estimate of $\dot{G}$ can be considered as a good rough approximation for all Galactic binaries detectable by LISA, even for high mass, high frequency binaries where the most reliable results are obtained using a chirping waveform as the frequency evolution cannot be ignored. We will now see that this is not the case for the parameter space of DECIGO.

\subsection{Chirping DECIGO sources}

For DECIGO, the right hand panel of Fig.~\ref{fig:contourGdot} shows that the quasi-monochromatic approximation breaks down in the same region where the Fisher matrix 
predicts interesting constraints on $\dot G$. 
We show that competitive constraints can indeed be achieved with these sources.

We perform a series of MCMC analyses with the same techniques described in the previous subsection in the area to the right of the red line in the right panel of Fig.~\ref{fig:contourGdot}. We find that chirping waveform models applied to DECIGO binaries can achieve constraints down to $\dot G/G_0\lesssim 10^{-11}\text{yr}^{-1}$ with the most favourable binaries in the sampled parameter space.

Even though the monochromatic approximation breaks down at higher frequencies and masses, we see that the quasi-monochromatic analytical Fisher matrix always predicts the true constraints within approximately an order of magnitude. 
Our quasi-monochromatic Fisher matrix analyis is more optimistic in the top right corner of parameter space, while it perfectly matches the MCMC chirping-waveform results close to the red line, where the quasi-monochromatic approximation starts to be valid.
This is expected, as in the delimiting region the chirping waveform model effectively resembles the Taylor-expanded model.
We also note that, while usually higher masses and higher frequencies produce better constraints, that is not always the case: for example, binaries with a chirp mass of $\log_{10}(\mathcal{M}_c/M_{\odot})= 1.4$ and a starting frequency of $\log_{10}(f_0/\text{Hz}) = -1.1 $ perform worse than binaries with a chirp mass of $\log_{10}(\mathcal{M}_c/M_{\odot})=1$ and a starting frequency of $\log_{10}(f_0/\text{Hz})= -1.4$. This is due to the fact that higher-mass, higher-frequency binaries spend less time inside the DECIGO frequency band since they quickly chirp out of band. This provides an effectively lower SNR with respect to the quasi-monochrmatic analysis which assumes that the binary is observed for the full duration of the DECIGO mission (4.5 yr).
Combining this effect with the usual trend dictating that higher masses and higher frequency provides better constraints, we find a sweet line a little below the top right corner of the right panel of Fig. \ref{fig:contourGdot} where DECIGO constraints on $\dot{G}$ will be the most stringent.
Such a region corresponds to binaries whose time to coalescence at the start of observations matches the total time duration of observations, i.e.~to the higher frequency, higher mass binaries that can be observed the longest.
As expected the bounds obtained in this region, as shown by the darker blue points along the white dashed line in the right panel of Fig.~\ref{fig:contourGdot}, represent in fact the best estimates we obtain in this paper.

\section{Discussion and conclusions}
\label{sec:discussion}

In this work, we forecast how well space-borne GW detectors could constrain the time evolution of the gravitational constant, $G$, with low-mass binaries.
The bounds we forecast are in some sense guaranteed, since sources like double WDs, binary NSs and binary stellar-mass BHs have already been observed either by EM surveys (DWDs) or by ground-based GW detectors (binary BHs and NSs).
This is not the case for other GW sources that can be used to forecast similar constraints on $\dot{G}$, such as massive BH binaries or EMRIs \cite{yunes2009post}.

According to our results, LISA will achieve the best constraints if our Galaxy hosts just one stellar-mass BH binary emitting at the upper end of LISA's frequency sensitivity range, for which we estimate to reach bounds of the order of $\dot G/G_0 \lesssim 10^{-11}\text{yr}^{-1}$. Recent simulations predict that LISA will detect tens to hundreds of binary BHs and NSs in the Milky Way~\cite{Wagg:2021cst}, some of which might fall in the most constraining region of parameter space. 
Lower-mass galactic binaries, in particular DWDs which LISA will detected in the tens of thousands, provide weaker constraints. {The most promising currently known DWD emitting GWs in the LISA band (usually referred to as verification binary) will give a bound $\dot G/G_0 \lesssim 10^{-4}\text{yr}^{-1}$. If, however, LISA detects at least one DWD at higher frequency, as predicted by population synthesis studies, then the bound is brought down to $\dot G/G_0 \lesssim 10^{-6}\text{yr}^{-1}$.}

The quasi-monochromatic assumption is quite restrictive for the parameter space covered by deci-Hz detectors. The region in which the assumption is valid does not yield competitive constraints for stellar-mass binaries detected with DECIGO at cosmological distances, the main target population of deci-Hz detectors.
For this reason, we have decided to include forecasts based on chirping waveforms within the stationary-phase approximation, {analogously to what was done} in Ref.~\cite{yunes2009post}. We find that indeed these give the best constraints overall in this paper, namely $\dot G/G_0 \lesssim 10^{-11}\text{yr}^{-1}$, for binaries at $D_L \sim 10$ Mpc which merge towards the end of the mission duration (i.e.~for which $\tau \simeq T_{\rm obs}$, with $\tau$ the time to coalescence).

The constraints forecast in this paper complement other analyses in the literature.
The analytical approach developed here is comparable to the one used to find constraints on $\dot{G}$ with pulsar timing, since they both target deviations in the GW emission at low-frequency and at Galactic distances. Although pulsar timing constraints are already surpassing the expectations from LISA~\cite{Kaspi:1994hp}, this spaceborne detector expected to fly in the 2030s will have access to GW sources all over the Milky Way and will test whether $G_0$ is indeed constant at different Galactic locations, with a method complementary to EM observations.
Note that at galactic distances only pulsar timing and GWs are known to give competitive constraints on $\dot{G}$ (cf.~Fig.~\ref{fig:LiteratureConstraints}).
The situation is different at cosmological distances, where {competitive constraints can be achieved with} other binary sources detectable by spaceborne GW detectors (e.g.~SOBH, EMRIs, or SMBHs) or Earth-based GW detectors (SOBHs or NSs).
At those distances, deci-Hz detectors such as DECIGO are expected to provide constraints comparable to (if not better than) the ones forecast with other GW sources.
Multi-band GW sources, detectable {by spaceborne and then}
Earth-based interferometers, are expected to provide even more stringent constraints, as recent analyses combining LISA and third-generation Earth-based detectors suggest~\cite{Perkins:2020tra}.
Such multi-band analyses are outside the scope of our present work, and are left for future considerations. 

Our results fall short of existing constraints obtained with very different methods and at very different distances. 
Solar System tests, for instance, already constrain $\dot G/G_0 \lesssim 10^{-14}\text{yr}^{-1}$~\cite{Genova}. Although orders of magnitude better than achievable with GW observations, this constraint is only valid locally, and obtained in a very different environment than the Galactic and cosmological ones probed by GWs.
Cosmological measurements from the early universe are also obtained in a completely different environment and with very different techniques.
Moreover, cosmological constraints are sensitive to the \emph{global} change in the value of $G(t)$ from the early universe to today, rather than the local time-derivative of $G(t)$ at the time of GW emission.
This is also true for tests of the running of $G$ based on NS masses~\cite{Vijaykumar:2020nzc}, which probe similar cosmological distances compared to binary coalescences, but can only measure the global variation of the value of $G$ from the time of merger to today.
GW inspiralling binaries, on the other hand, offer a method to test localised time-variation of $G$ at any Galactic and cosmological distance, up to Gpc scales.

We stress that to reach our results we have performed an extensive study of the parameter space of quasi-monochromatic binaries, for both milli-Hz and deci-Hz sources, and we have confirmed the results with targeted MCMC analyses.
Such an approach allowed us to identify the most constraining region in parameter space, and consequently to identify the best GW source population to use to search for variations in $G$.

While we have focused here on the upcoming LISA mission and the proposed DECIGO detector, our study could be easily extended to other low-frequency detector designs, such as TianQin~\cite{TianQin:2015yph}, $\mu$Ares \cite{Sesana:2019vho}, ALIA \cite{Baker:2019pnp} or other deci-Hz designs \cite{Sedda:2021yhn}.
The analytical framework outlined in Sec.~\ref{sec:analyticalapproach} could also be applied to explore a wide range of effects that might influence binary inspirals at large separations. Further corrections to the phase of the GWs at $-4$PN are predicted to arise from binaries' peculiar accelerations \cite{Tamanini_2020, Inayoshi_2017,Bonvin_2017,Wong_2019,Randall_2019}, matter accretion \cite{Sberna_2021, Kremer_2017,Caputo_2020,Cardoso_2020}, dynamical friction \cite{Cardoso_2020} and enhanced black hole evaporation due to extra dimensions \cite{Emparan_2003, Berti_2018}. These effects are all degenerate to a first approximation, so any of these effects can only be detected individually if we can assume it dominates over the others (depending, e.g., on the astrophysical configuration of the binary), or if it can be discerned using a population of binaries for which the same effect may be different for different binaries, as expected for example for peculiar accelerations.
Note also that here we find that the binaries yielding the best constraints are the one merging around the time GW observations stops, namely for which $\tau = T_{\rm obs}$.
This is in agreement with what found in previous work targeting similar -4PN effects~\cite{Inayoshi_2017,Tamanini_2020}.

To conclude our results show that stellar-mass GW binaries detectable with future spaceborne detectors can offer new, complementary and possibly competitive constraints on the local time-evolution of Newton's constant at distances ranging from Galactic to cosmological scales.
Testing the constancy of $G$, and more in general the validity of general relativity, at different scales and with different methods will definitely help us better understand the behaviour of the gravitational interation in our universe.

\begin{acknowledgements}
We thank Sylvain Marsat for insightful discussions and access to the LISAbeta code, and Valeriya Korol for providing the data behind the estimates with the full DWD population reported in Sec.~\ref{subsec:LISA_analytic}.
A.A.~acknowledges support from NSF Grants No. PHY-1912550, AST-2006538, PHY-090003 and PHY-20043, and NASA Grants No.~17-ATP17-0225, 19-ATP19-0051 and 20-LPS20-0011.
N.T.~acknowledges support form the French space agency CNES in the framework of LISA and by an ANR Tremplin ERC grant (ANR-20-ERC9-0006-01).
\end{acknowledgements}

\bibliographystyle{IEEEtran}
\bibliography{AMG_paperNotes}

\end{document}